\documentclass[aps,prl,twocolumn,superscriptaddress,showpacs]{revtex4-1}

\usepackage{graphicx}
\usepackage[T1]{fontenc}
\usepackage{multirow}
\usepackage{tabularx}
\usepackage{xcolor}
\usepackage{threeparttable}

\usepackage{bm}
\begin{document}

\title{How Robust is the $\emph{N}$ = 34 Subshell Closure? First Spectroscopy of $^{52}$Ar}

\newcommand{\atexas}{      \affiliation{Texas A$\&$M University-Commerce, PO Box 3011, Commerce, Texas 75429, USA}}
\newcommand{\akth}{        \affiliation{Department of Physics, Royal Institute of Technology, SE-10691 Stockholm, Sweden}}
\newcommand{\aatomki}{     \affiliation{MTA Atomki, P.O. Box 51, Debrecen H-4001, Hungary}}
\newcommand{\abeijing}{    \affiliation{State Key Laboratory of Nuclear Physics and Technology, Peking University, Beijing 100871, P.R. China}}
\newcommand{\acaen}{       \affiliation{LPC Caen, ENSICAEN, Universit\'e de Caen, CNRS/IN2P3, F-14050 Caen, France}}
\newcommand{\acea}{        \affiliation{D\'{e}partement de Physique Nucl\'{e}aire, IRFU, CEA, Universit\'e Paris-Saclay, F-91191 Gif-sur-Yvette, France}}
\newcommand{\acns}{        \affiliation{Center for Nuclear Study, University of Tokyo, RIKEN campus, Wako, Saitama 351-0198, Japan}}
\newcommand{\aewha}{       \affiliation{Ewha Womans University, Seoul, Korea}}
\newcommand{\agsi}{        \affiliation{GSI Helmoltzzentrum f\"ur Schwerionenforschung GmbH, 64291 Darmstadt, Germany}}
\newcommand{\aemiegsi}{        \affiliation{ExtreMe Matter Institute EMMI, GSI Helmholtzzentrum f\"ur Schwerionenforschung GmbH, 64291 Darmstadt, Germany}}
\newcommand{\croatia}{      \affiliation{Ru{\dj}er Bo\v{s}kovi\'{c} Institute, Bijeni\v{c}ka cesta 54,10000 Zagreb, Croatia}}
\newcommand{\ahku}{        \affiliation{Department of Physics, The University of Hong Kong, Pokfulam Road, Hong Kong}}
\newcommand{\ainst}{       \affiliation{Institute for Nuclear Science $\&$ Technology, VINATOM, P.O.Box 5T-160, Nghia Do, Hanoi, Vietnam}}
\newcommand{\aipno}{       \affiliation{Institut de Physique Nucl\'eaire, CNRS-IN2P3, Universit\'e Paris-Sud, Universit\'e Paris-Saclay, 91406 Orsay Cedex, France}}
\newcommand{\akoeln}{      \affiliation{Institut f\"ur Kernphysik, Universit\"at zu K\"oln, D-50937 Cologne, Germany}}
\newcommand{\alanzhou}{    \affiliation{Institute of Modern Physics, Chinese Academy of Sciences, Lanzhou 730000, China}}
\newcommand{\amadrid}{     \affiliation{Instituto de Estructura de la Materia, CSIC, E-28006 Madrid, Spain}}
\newcommand{\aorsay}{      \affiliation{CSNSM, CNRS/IN2P3, Universit\'e Paris-Sud, F-91405 Orsay Campus, France}}
\newcommand{\aoslo}{       \affiliation{Department of Physics, University of Oslo, N-0316 Oslo, Norway}}
\newcommand{\ariken}{      \affiliation{RIKEN Nishina Center, 2-1 Hirosawa, Wako, Saitama 351-0198, Japan}}
\newcommand{\arikkyo}{     \affiliation{Department of Physics, Rikkyo University, 3-34-1 Nishi-Ikebukuro, Toshima, Tokyo 172-8501, Japan}}
\newcommand{\atitech}{     \affiliation{Department of Physics, Tokyo Institute of Technology, 2-12-1 O-Okayama, Meguro, Tokyo, 152-8551, Japan}}
\newcommand{\atohoku}{     \affiliation{Department of Physics, Tohoku University, Sendai 980-8578, Japan}}
\newcommand{\atudarmstadt}{\affiliation{Institut f\"ur Kernphysik, Technische Universit\"at Darmstadt, 64289 Darmstadt, Germany}}
\newcommand{\aunal}{       \affiliation{Universidad Nacional de Colombia, Sede Bogota, Facultad de Ciencias, Departamento de F\'isica, 111321, Bogot\'a, Colombia}}
\newcommand{\aut}{         \affiliation{Department of Physics, University of Tokyo, 7-3-1 Hongo, Bunkyo, Tokyo 113-0033, Japan}}
\newcommand{\aoak}{         \affiliation{Physics Division, Oak Ridge National Laboratory, Oak Ridge, Tennessee 37831, USA}}
\newcommand{\anccs}{         \affiliation{National Center for Computational Sciences, Oak Ridge National Laboratory, Oak Ridge, Tennessee 37831, USA}}
\newcommand{\atennessee}{         \affiliation{Department of Physics and Astronomy, University of Tennessee, Knoxville, Tennessee 37996, USA}}
\newcommand{\atriumf}{         \affiliation{TRIUMF 4004 Wesbrook Mall, Vancouver, British Columbia V6T 2A3, Canada}}
\newcommand{\amaxplank}{         \affiliation{Max-Planck-Institut f\"ur Kernphysik, Saupfercheckweg 1, 69117 Heidelberg, Germany}}

\author{H.~N.~Liu}
\email{hliu@ikp.tu-darmstadt.de}
\acea \atudarmstadt \akth
\author{A.~Obertelli} \atudarmstadt \acea \ariken
\author{P.~Doornenbal} \ariken
\author{C.~A. Bertulani} \atexas \atudarmstadt
\author{G.~Hagen} \aoak \atennessee
\author{J.~D.~Holt} \atriumf
\author{G.~R.~Jansen} \anccs \aoak
\author{T.~D.~Morris} \aoak \atennessee 
\author{A.~Schwenk} \atudarmstadt \aemiegsi \amaxplank
\author{R.~ Stroberg} \atriumf
\author{N.~Achouri}\aipno
\author{H.~Baba} \ariken
\author{F.~Browne}\ariken
\author{D.~Calvet} \acea
\author{F.~Ch\^ateau} \acea
\author{S.~Chen} \abeijing \ariken \ahku
\author{N.~Chiga}\ariken 
\author{A.~Corsi} \acea 
\author{M.~L.~Cort\'es} \ariken
\author{A.~Delbart} \acea
\author{J.-M.~Gheller} \acea
\author{A.~Giganon}\acea 
\author{A.~Gillibert} \acea
\author{C.~Hilaire} \acea
\author{T.~Isobe} \ariken 
\author{T.~Kobayashi}\atohoku
\author{Y.~Kubota} \ariken \acns
\author{V.~Lapoux} \acea
\author{T.~Motobayashi} \ariken 
\author{I.~Murray}\aipno \ariken
\author{H.~Otsu} \ariken
\author{V.~Panin}\ariken
\author{N.~Paul} \acea
\author{W.~Rodriguez}\aunal \ariken
\author{H.~Sakurai} \ariken \aut
\author{M.~Sasano} \ariken
\author{D.~Steppenbeck}\ariken
\author{L.~Stuhl}\acns
\author{Y.~L.~Sun}\acea \atudarmstadt
\author{Y.~Togano}\arikkyo  
\author{T.~Uesaka} \ariken
\author{K.~Wimmer}\aut
\author{K.~Yoneda} \ariken 
\author{O.~Aktas}\akth
\author{T.~Aumann}\atudarmstadt  
\author{L.~X.~Chung} \ainst 
\author{F.~Flavigny}\aipno
\author{S.~Franchoo}\aipno
\author{I.~Ga\v{s}pari\'{c}}\croatia \ariken   
\author{R.~-B.~Gerst}\akoeln
\author{J.~Gibelin}\acaen
\author{K.~I.~Hahn} \aewha
\author{D.~Kim} \aewha \ariken
\author{T.~Koiwai} \aut  
\author{Y.~Kondo}\atitech   
\author{P.~Koseoglou}\atudarmstadt \agsi
\author{J.~Lee} \ahku
\author{C.~Lehr}\atudarmstadt 
\author{B.~D.~Linh}\ainst 
\author{T.~Lokotko}\ahku
\author{M.~MacCormick}\aipno
\author{K.~Moschner}\akoeln
\author{T.~Nakamura}\atitech 
\author{S.~Y.~Park} \aewha \ariken
\author{D.~Rossi}\atudarmstadt   
\author{E.~Sahin} \aoslo
\author{D.~Sohler}\aatomki  
\author{P.-A.~S\"oderstr\"om} \atudarmstadt
\author{S.~Takeuchi} \atitech 
\author{H.~T\"ornqvist}\agsi 
\author{V.~Vaquero}\amadrid 
\author{V.~Wagner}\atudarmstadt 
\author{S.~Wang}\alanzhou
\author{V.~Werner}\atudarmstadt
\author{X.~Xu} \ahku
\author{H.~Yamada}\atitech 
\author{D.~Yan} \alanzhou
\author{Z.~Yang} \ariken 
\author{M.~Yasuda}\atitech  
\author{L.~Zanetti}\atudarmstadt 

\date{\today}

\begin{abstract}

  The first $\gamma$-ray spectroscopy of $^{52}$Ar, with the neutron number $\emph{N}$ = 34, was measured using the $^{53}$K(\emph{p},2\emph{p}) one-proton removal reaction at $\sim$210 MeV/u at the RIBF facility. The 2$^{+}_{1}$ excitation energy is found at 1656(18) keV, the highest among the Ar isotopes with $\emph{N}$ $>$ 20. This result is the first experimental signature of the persistence of the $\emph{N}$ = 34 subshell closure beyond $^{54}$Ca, i.e., below the magic proton number $\emph{Z}$ = 20. Shell-model calculations with phenomenological and chiral-effective-field-theory interactions both reproduce the measured 2$^{+}_{1}$ systematics of neutron-rich Ar isotopes, and support a $\emph{N}$ = 34 subshell closure in $^{52}$Ar.

\end{abstract}

\pacs{xxxxxxxx}

\maketitle

In the shell-model description of atomic nuclei, magic numbers of nucleons correspond to fully occupied energy shells below the Fermi surface \cite{may55}, and present the backbone of our understanding of nuclei. Scientific advances over the past decades have shown that the sequence of magic numbers established for stable nuclei is not universal across the nuclear landscape \cite{sorlin08}. A few prominent examples are the breakdown of the conventional $\emph{N}$ = 20, 28 magic numbers \cite{thibault75,mue84,ba07} and the emergence of a new $\emph{N}$ = 16 magic number \cite{ozawa00,tshoo12} in neutron-rich nuclei. Considerable efforts have been spent to unfold the driving forces behind such shell evolution \cite{tal60,ots05,zuker03,ots10}.

For rare isotopes, the first 2$^{+}$ excitation energy [$\emph{E}$(2$^{+}_{1}$)] in even-even nuclei is often the first observable accessible to experiment to characterize shell effects. In a simplified shell model picture, a high $\emph{E}$(2$^{+}_{1}$) is interpreted as resulting from nucleon excitations across a large shell gap \cite{bohr88}.

Recently, neutron-rich $\emph{pf}$-shell nuclei have received much attention on both experimental and theoretical fronts with the possible appearance of new subshell closures at $\emph{N}$ = 32 and 34.
A sizable $\emph{N}$ = 32 subshell closure has been reported in the region from Ar to Cr isotopes based on $\emph{E}$(2$^{+}_{1}$) \cite{huck85,jan02,pri01,step15}, reduced transition probabilities $\emph{B}$($\emph{E}$2; 0$_{1}^{+}$$\rightarrow$2$_{1}^{+}$) \cite{din05,bur05}, and mass \cite{wie13,ro15,xu15} measurements, although some ambiguity remains due to the newly measured large charge radii of $^{49,51,52}$Ca \cite{rui16}, masses of $^{51-55}$Ti \cite{lei18}, and the low $\emph{E}$(2$^{+}_{1}$) in $^{50}$Ar \cite{step15}.
On the other hand, the $\emph{N}$ = 34 subshell closure has been so far suggested only in $^{54}$Ca \cite{step13,mi18}.
In the Ti and Cr isotopes, the systematics of $\emph{E}$(2$^{+}_{1}$) \cite{su13,zhu06} and $\emph{B}$($\emph{E}$2; 0$_{1}^{+}$$\rightarrow$2$_{1}^{+}$) \cite{din05, bur05} show no local maximum and minimum at $\emph{N}$ = 34. The measured low-lying structure of $^{55}$Sc \cite{step17} indicated a rapid weakening of the $\emph{N}$ = 34 subshell closure in Z $>$ 20 nuclei.
The measured $\emph{E}$(2$_{1}^{+}$) of $^{54}$Ca was 2043(19) keV, $\sim$0.5 MeV lower than $^{52}$Ca \cite{step13}. Despite this lower $\emph{E}$(2$^{+}_{1}$), $^{54}$Ca was concluded to be a doubly magic nucleus from a phenomenological shell-model interpretation \cite{step13}, whereas $\emph{ab initio}$ coupled-cluster calculations indicated a weak $\emph{N}$ = 34 subshell closure \cite{hagen12}. Very recently, the mass measurements of $^{55-57}$Ca \cite{mi18} confirmed the $\emph{N}$ = 34 subshell closure in $^{54}$Ca.
Until now, it is still an open question how the $\emph{N}$ = 34 subshell evolves below $\emph{Z}$ = 20 towards more neutron-rich systems, such as $^{52}$Ar.

The heaviest Ar isotope with known spectroscopic information so far is $^{50}$Ar \cite{step15}. Phenomenological shell-model calculations \cite{step15,utsu14} reproducing the available $\emph{E}$(2$^{+}_{1}$) data for neutron-rich Ar isotopes predict a relatively high-lying 2$^{+}_{1}$ state in $^{52}$Ar, and suggest that the $\emph{N}$ = 34 subshell closure in $^{52}$Ar is stronger than the one reported for $^{54}$Ca. In the present Letter, we report on the first spectroscopy of $^{52}$Ar, the most neutron-rich even-even $\emph{N}$ = 34 isotone accessible today and possibly for the next decades. A clear enhancement of $\emph{E}$(2$^{+}_{1}$) at $\emph{N}$ = 34 is found, supporting the persistence of the $\emph{N}$ = 34 magic number in $\emph{Z}$ $<$ 20 nuclei.


\begin{figure}[t!]
  \centering
    \includegraphics[width=8.6cm]{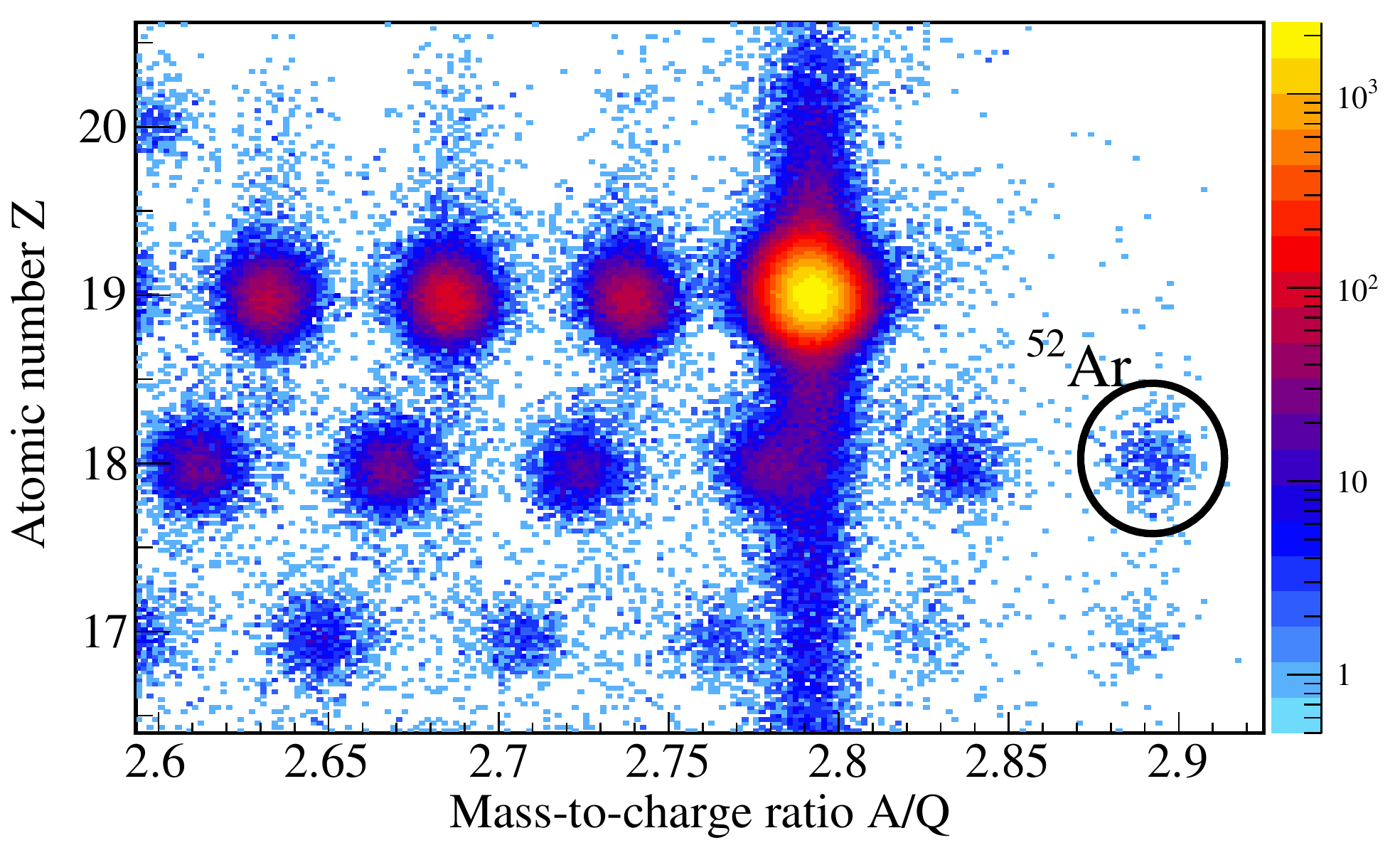}
    \caption{(Color online) Particle identification plot of reaction residues with the selection of $^{53}$K identified at BigRIPS.
  }\label{fig:1}
\end{figure}


The experiment was performed at the Radioactive Isotope Beam Factory operated by the RIKEN Nishina Center and the Center for Nuclear Study of the University of Tokyo. Radioactive nuclei were produced by fragmentation of a 345 MeV/u $^{70}$Zn primary beam with an average beam intensity of 240 pnA on a 10-mm-thick rotating Be target. The secondary cocktail beam, magnetically centered on $^{53}$K, was identified using the $\emph{B}$$\rho$-$\Delta$$\emph{E}$-TOF method \cite{fukuda13} in the BigRIPS two-stage fragment separator \cite{ku12}. The average intensity and purity of the $^{53}$K beam were 1.0 particle per second and 0.53\%, respectively.

The secondary beam impinged on a 151(1)-mm-thick liquid hydrogen (LH$_{2}$) target operated at 18.14 K with a density of 73 kg/m$^{3}$ to induce one-proton knockout reactions. Two multiwire drift chambers (MDCs) \cite{ko13}, located upstream of the LH$_{2}$ target, were used to measure the trajectories of the incoming projectiles. The kinetic energy of $^{53}$K at the entrance of the target was $\sim$245 MeV/u. Its energy loss in the LH$_{2}$ target was approximately 70 MeV/u. The LH$_{2}$ target was surrounded by a 300-mm-long time projection chamber (TPC), constituting the MINOS device \cite{ob14}. The tracks of outgoing protons were recorded by the TPC to reconstruct reaction vertices \cite{san18}. The measured efficiency to detect at least one of the two protons is 92(3)\% with an estimated vertex resolution of 4 mm (FWHM) \cite{san18}. Knowledge of reaction vertices allowed precise Doppler correction of de-excitation $\gamma$ rays measured by the DALI2+ \cite{ta14} $\gamma$-ray spectrometer, which surrounded the MINOS device.

DALI2+ consisted of 226 NaI(Tl) crystals covering polar angles from 15$^\circ$ to 118$^\circ$ with respect to the center of the LH$_{2}$ target. Thresholds of NaI(Tl) crystals were set to about 50 keV. Addback was applied when the centers of hit detectors were less than 20 cm apart. For 1 MeV $\gamma$ rays emitted from nuclei moving at 60\% of the velocity of light, the photopeak efficiency and energy resolution with addback were 30\% and 11\% (FWHM), respectively. DALI2+ was calibrated using $^{133}$Ba, $^{137}$Cs, $^{60}$Co, and $^{88}$Y sources yielding a calibration error of 4 keV and a good linearity from 356 to 1836~keV. 

Downstream from the LH$_{2}$ target, reaction residues were transported to the SAMURAI spectrometer \cite{ko13} and identified with the $\emph{B}$$\rho$-$\Delta$$\emph{E}$-TOF method. The $\emph{B}$$\rho$ of charged fragments passing through the SAMURAI magnet with a central magnetic field of 2.7 T was reconstructed using two MDCs placed upstream and downstream the magnet \cite{ko13}. The $\Delta$$\emph{E}$ and TOF information were provided by a 24-element plastic scintillator hodoscope. Figure 1 shows the particle identification of fragments with the selection of $^{53}$K identified at BigRIPS. For Ar isotopes, a 6.6$\sigma$ separation in $\emph{Z}$ and a 9.1$\sigma$ separation in $\emph{A}$ were achieved. Over the data taking of seven days, 438 counts of $^{52}$Ar were accumulated from $^{53}$K($\emph{p}$, 2$\emph{p}$)$^{52}$Ar reactions, in which the kinematics of protons measured by MINOS supported a quasi-free scattering reaction mechanism. The reaction loss of $^{53}$K in materials along the beam and fragment trajectories was determined by measuring the unreacted $^{53}$K. The measured inclusive cross section was 1.9(1) mb.

The Doppler-shift corrected $\gamma$-ray energy spectrum of $^{52}$Ar is presented in Fig. 2. A clear peak is present in the range of 1500-1800 keV, while three structures are visible in the range of 600-900, 1000-1300 and 2000-2500 keV. In order to quantify the significance level of these peaks, we performed the likelihood ratio test by fitting the spectrum of $^{52}$Ar using the GEANT4 \cite{ago03} simulated response functions on top of a double-exponential background.
Given the low statistics of the $\gamma$-ray spectrum of $^{52}$Ar, a Poisson distribution was adopted to describe the fluctuations of each bin, and the double-exponential background line shape was extracted from the $^{51}$K($\emph{p}$,2$\emph{p}$)$^{50}$Ar reaction, leaving the magnitude of the background as a free parameter in the fitting. 
To estimate the systematic uncertainties caused by this background assumption, the spectrum of $^{52}$Ar was also fitted using a free double-exponential background, as well as background line shapes extracted from $^{54}$Ca($\emph{p}$,$\emph{pn}$)$^{53}$Ca and $^{55}$V($\emph{p}$,2$\emph{pn}$)$^{53}$Ca reactions. Note that $^{53}$Ca has similar transitions and neutron separation energy (\emph{S$\rm{_{n}}$}) as $^{52}$Ar.
As a result, a significance level of 5$\sigma$ was obtained for the 1656(18)~keV transition. The 2295(39)~keV $\gamma$ line was found to have a significance of 3$\sigma$, while the other two structures in the range of 600-900 and 1000-1300 keV both had a significance level of less than 1$\sigma$ and are therefore not considered in the following analysis.
Note that errors of the deduced $\gamma$-ray energies shown above include both statistical and systematic uncertainties. The former dominate and the latter mainly originate from the energy calibration uncertainty. Lifetime ($\tau_{\gamma}$) effects of the excited states on the deduced $\gamma$-ray energies are negligible, since Raman's global systematics \cite{raman01} suggests $\tau_{\gamma}$ $<$ 2 ps for the observed two states.


\begin{figure}[t!]
  \centering
    \includegraphics[width=8.6cm]{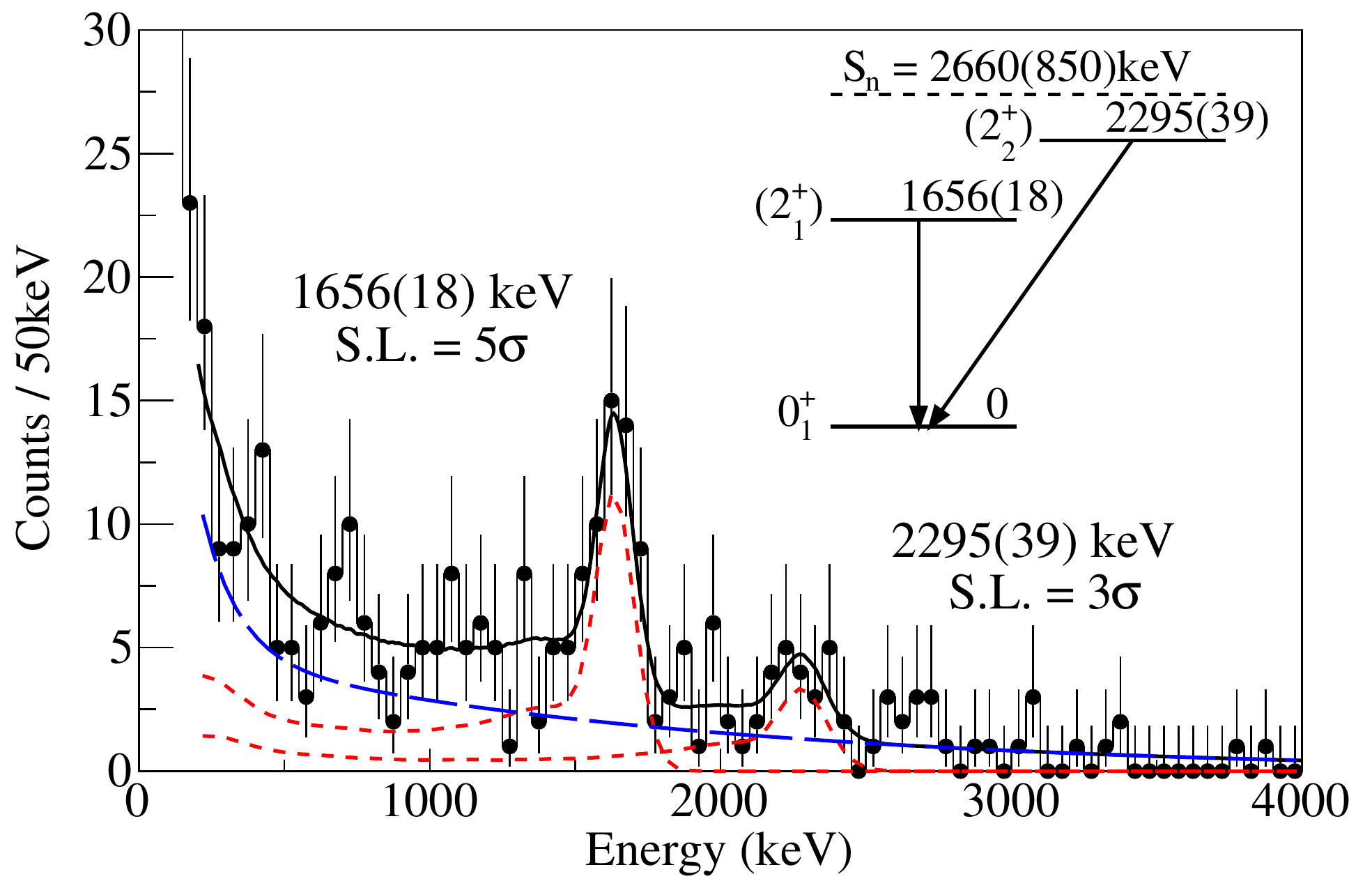}\\
    \caption{(Color online)  Doppler-shift corrected $\gamma$-ray energy spectrum of $^{52}$Ar following the $^{53}$K($\emph{p}$,2$\emph{p}$) reaction. The fit function to the spectrum (black solid line) includes simulated response functions for the observed transitions (red dotted line) and a double-exponential background (blue dashed line).  
 The significance level (S.L.) is given for the observed transitions. The insert shows the deduced experimental level scheme.
  }\label{fig:2}
\end{figure}


Based on measured $\gamma$-ray intensities, the 1656~keV transition is attributed to a direct decay to the ground state. The low statistics do not allow to conclude any (non) coincidence between the 1656 and 2295~keV transitions from $\gamma$--$\gamma$ correlations. However, the cascade scenario is very unlikely due to the expected low \emph{S$\rm{_{n}}$} of $^{52}$Ar that is more exotic than $^{54}$Ca.
The measured \emph{S$\rm{_{n}}$} of $^{54}$Ca is 3840(70) keV \cite{wie13}. The 2016 Atomic Mass Evaluation \cite{wang17} gives an estimated \emph{S$\rm{_{n}}$} of 2660(850) keV for $^{52}$Ar, and excludes the coincidence scenario. The proposed energy level scheme of $^{52}$Ar is presented in the inset of Fig. 2. The measured partial cross sections to the 1656 and 2295 keV states are 0.9(2) and 0.4(1)~mb, respectively. 
Assuming no population to other excited states, the cross section to the ground state is deduced to be 0.7(3) mb via subtraction from the inclusive cross section. The quoted uncertainties are dominated by statistical errors, while systematic uncertainties mainly arise from the estimation of MINOS efficiency and the background assumption. All experimental results are summarized in Table \ref{ttexp}. The 1656~keV state with the higher population is assigned as 2$_{1}^{+}$. The 2295~keV state decaying directly to the ground state is assigned as 2$_{2}^{+}$. Further discussions about these spin-parity assignments are given later.
\raggedbottom

Figure 3 displays the measured $\emph{E}$(2$_{1}^{+}$) in $^{52}$Ar alongside values for lighter Ar isotopes \cite{nndc}. Notably, the measured $\emph{E}$(2$_{1}^{+}$) = 1656(18) keV in $^{52}$Ar is the highest among the Ar isotopes with $\emph{N}$~$>$~20. It is larger than the $\emph{E}$(2$_{1}^{+}$)~=~1554(1)~keV~\cite{nowak16} in $^{46}$Ar which reflects the conventional $\emph{N}$ = 28 shell closure. Moreover, the measured systematics of $\emph{E}$(2$_{1}^{+}$) along the Ar isotopic chain is characterized by a pronounced enhancement at $\emph{N}$ = 34 relative to its $\emph{N}$ = 32 even-even neighbor, unlike the trend observed for Ca, Ti and Cr isotopes in which a decrease is seen from $\emph{N}$ = 32 to 34. Our results offer the first experimental signature of the $\emph{N}$ = 34 subshell closure in $^{52}$Ar.

To gain further insight into the structure of $^{52}$Ar, we compare the results to state-of-the-art nuclear structure calculations. Here, two advanced $\emph{ab initio}$ approaches are adopted: the valence-space in-medium similarity renormalization group (VS-IMSRG) \cite{tsu12,str16,her16,str17} (for calculational details, see in particular Refs. \cite{str17, sim17}) and coupled-cluster theory \cite{bar07,hagen14}, employing two sets of two- ($\emph{NN}$) and three-nucleon (3$\emph{N}$) interactions derived from chiral effective field theory \cite{epe09,mac11}: 1.8/2.0(EM) \cite{heb11,sim16,sim17} and N$^{2}$LO$\rm_{sat}$ \cite{eks15}.  The coupled-cluster method is well suited for closed (sub-)shell nuclei and their neighbors. By employing a double-charge exchange equation-of-motion (DCE-EOM) coupled-cluster technique, the $\emph{E}$(2$_{1}^{+}$) in $^{40,48,52}$Ar are obtained from generalized excitations of the ground states of the closed (sub-)shell nuclei $^{40,48,52}$Ca, while the $\emph{E}$(2$_{1}^{+}$) in $^{44}$Ar is obtained from excitations of the $^{44}$S ground state, respectively. The $\emph{E}$(2$_{1}^{+}$) in $^{46,50,52}$Ar are also computed using the two-particle-removed equation-of-motion (2PR-EOM) coupled-cluster method~\cite{jansen2011}. The measured $\emph{E}$(2$_{1}^{+}$) in $^{52}$Ar offers a rather unique case to compare these two coupled-cluster methods. 

In this work, we employ the DCE-EOM coupled-cluster calculations with particle-hole excitations truncated at the singles, doubles, and approximate triples level (CCSDT-3) \cite{watts1996}, while the 2PR-EOM coupled-cluster calculations are truncated at the three-hole-one-particle excitation level using CCSD and CCSDT-3 for the ground states of $^{48,52,54}$Ca. Theoretical uncertainties in coupled-cluster calculations are estimated by comparing results with and without triples excitations.
In addition, we also compare our results to large-scale shell model calculations with the SDPF-U \cite{nowacki09} and SDPF-MU~\cite{utsu12} interactions. Note that the original SDPF-MU Hamiltonian was modified using recent experimental data on exotic Ca \cite{step13} and K \cite{pap13} isotopes and details of the modifications are given in Ref. \cite{utsu14}. 

\begin{table}[t!]
\small
\caption{\label{ttexp} Experimental excitation energies ($\emph{E}\rm_{exp}$) and cross sections ($\sigma\rm_{exp}$) from the $^{53}$K($\emph{p}$,2$\emph{p}$)$^{52}$Ar reaction in comparison with theoretical calculations. Predicted excitation energies ($\emph{E}_{x}$), \emph{J$^{\pi}$}, and spectroscopic factors ($\emph{C}^{2}\emph{S}\rm_{th}$) associated with the removed protons from different orbits ($\emph{l}$$_{j}$) were obtained using the VS-IMSRG method predicting a $^{53}$K(3/2$^{+}$) ground state. Theoretical partial cross sections ($\sigma\rm_{th}$) were computed using the $\emph{C}^{2}\emph{S}\rm_{th}$ values and beam-energy-weighted average single-particle cross sections ($\langle\sigma\rm_{sp}\rangle$).}
\centering
\begin{threeparttable}
\begin{tabular*}{8.6cm}{@{\extracolsep{\fill}}cc|cccccc}
\hline
\hline
\multicolumn{2}{c|}{Experiment} & \multicolumn{6}{c}{Theory}\\

\hline
$\emph{E}\rm_{exp}$ & $\sigma\rm_{exp}$ & $\emph{E}_{x}$ & \emph{J$^{\pi}$} & $\emph{l}$$_{j}$ & $\emph{C}^{2}\emph{S}\rm_{th}$ & $\langle\sigma\rm_{sp}\rangle$ & $\sigma\rm_{th}$ \\
 (keV) &    (mb)             &    (keV)            &                                  &  &  & (mb)  & (mb)\\
\hline
0 &  0.7(3)\tnote{a} & 0 & 0$_{1}^{+}$ & $\emph{d}$$_{3/2}$  & 0.28 & 3.03 & 0.86 \\
\hline
1656(18)&  0.9(2) & 1849 & 2$_{1}^{+}$ & $\emph{s}$$_{1/2}$ & 0.10 & 0.92 & 1.13\\
& &      &            & $\emph{d}$$_{3/2}$ & 0.33 & 2.94 &  \\
& &      &            & $\emph{d}$$_{5/2}$ & 0.02 & 4.82 & \\
 \hline
& &  1974 & 0$_{2}^{+}$ & $\emph{d}$$_{3/2}$ & 0.01 & 2.93   & 0.04\\
 \hline
2295(39)  &  0.4(1) & 2367 & 2$_{2}^{+}$ &$\emph{s}$$_{1/2}$ & 0.13 & 0.92   & 0.30 \\
& &      &            & $\emph{d}$$_{3/2}$ & 0.05 & 2.91   & \\
& &      &            & $\emph{d}$$_{5/2}$ & 0.01 & 4.76   &\\
\hline
Inclusive & 1.9(1) &  &  &  &   & & 2.32 \\
\hline
\hline
\end{tabular*}
\begin{tablenotes}
\footnotesize
\item[a]{Deduced by assuming no population to other excited states except the 1656- and 2295-keV state as described in text.}
\end{tablenotes}
\end{threeparttable}
\end{table}


Theoretical ($\emph{p}$,2$\emph{p}$) cross sections to different final states of $^{52}$Ar are computed with spectroscopic factors calculated with the VS-IMSRG method using the 1.8/2.0 (EM) interaction and single-particle cross sections ($\sigma\rm_{sp}$) calculated using the Glauber theory as described in Ref. \cite{auman13}. The input of the $\sigma\rm_{sp}$ calculations are the nucleon-nucleon cross sections, using the parametrization from Ref. \cite{ber10}, and the nuclear ground-state densities deduced from a mean-field Hartree-Fock-Bogoliubov calculation using the SLy4 interaction. The involved single-particle states are calculated using a Woods-Saxon potential including the Coulomb and spin-orbit terms with parameters chosen to reproduce the proton separation energies. The range of the Woods-Saxon potential was taken as $R=r_0 (A-1)^{1/3}$ fm with $r_0 = 1.25$ fm, and the diffuseness is chosen as 0.65 fm. The strength of the spin-orbit potential is set to $-6$~MeV.
Since the reaction vertices were reconstructed with MINOS, the energy dependence of the cross section was considered by taking the average of $\sigma\rm_{sp}$ at different incident energies weighted by observed statistics ($\langle\sigma\rm_{sp}\rangle$). As an illustration, $\sigma\rm_{sp}$ for the removal of a $\emph{d}_{3/2}$ proton in $^{53}$K to the ground state of $^{52}$Ar varies from 2.38 mb at 180~MeV/u to 3.64~mb at 246~MeV/u. 
Table \ref{ttexp} lists theoretical results for states lying below the extrapolated \emph{S$\rm{_{n}}$} of $^{52}$Ar \cite{wang17}.
As shown in Table \ref{ttexp}, the measured cross sections to the 1656 and 2295 keV states in $^{52}$Ar agree well with the predictions for the 2$_{1}^{+}$ state at 1849~keV and the 2$_{2}^{+}$ state at 2367~keV, respectively.
The ratio of the experimental cross section to the theoretical prediction is in line with the systematic reduction factor reported from (\emph{e},\emph{e$^{\prime}$p}) measurements on stable targets \cite{lap93} and from ($\emph{p}$,2$\emph{p}$) reactions on oxygen isotopes \cite{atar18,kawase18}.
The good agreement between experiment and theory not only supports the spin-parity assignments, but also indicates that the VS-IMSRG approach with the 1.8/2.0 (EM) interaction provides a satisfactory description of the structure of $^{52}$Ar.

\begin{figure}[t!]
  \centering
    \includegraphics[width=8.68cm]{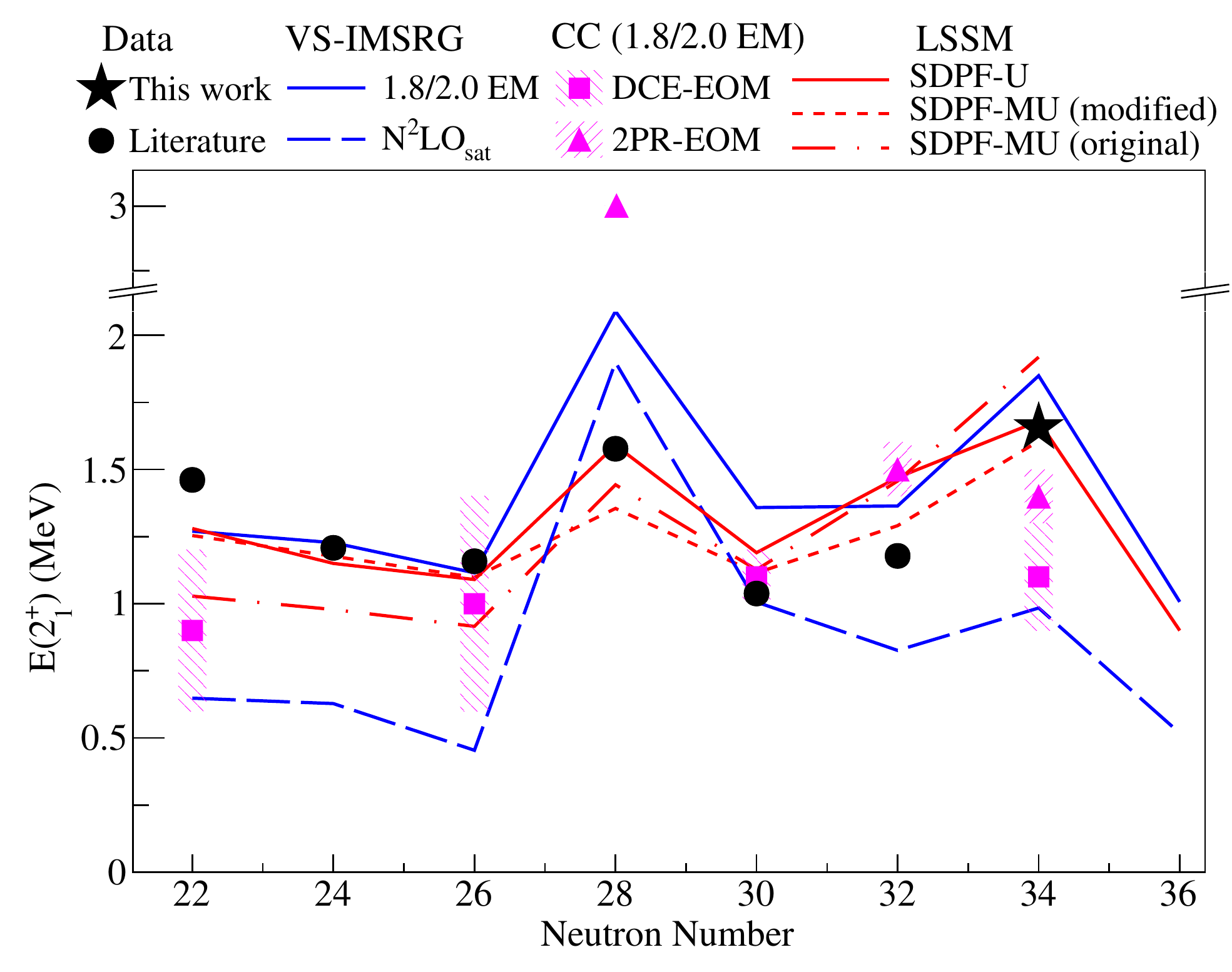}\\
    \caption{(Color online) Experimental 2$_{1}^{+}$ energies for even-even Ar isotopes compared with theory: VS-IMSRG with the chiral interaction 1.8/2.0(EM) \cite{heb11,sim16,sim17} and N$^{2}$LO$\rm_{sat}$ \cite{eks15}, coupled-cluster calculations (CC) using the DCE-EOM and 2PR-EOM methods with the 1.8/2.0(EM) interaction, and large-scale shell model (LSSM) calculations with the SDPF-U \cite{nowacki09} and the so-called \emph{original} \cite{utsu12} and \emph{modified} \cite{utsu14} SDPF-MU interactions. The hatched regions represent the theoretical uncertainties in coupled-cluster calculations. Experimental data are taken from Ref. \cite{nndc,nowak16} and this work. Note the broken y-axis scale between 2.2 and 2.7 MeV. 
  }\label{fig:3}
\end{figure}

We now discuss the systematics of $\emph{E}$(2$_{1}^{+}$) in Ar isotopes. As seen in Fig. 3, phenomenological shell-model calculations with the SDPF-U and SDPF-MU interactions as well as the VS-IMSRG calculations with the 1.8/2.0 (EM) interaction reproduce the steep rise of $\emph{E}$(2$_{1}^{+}$) from $^{50}$Ar to $^{52}$Ar.
The SDPF-U calculations and the modified SDPF-MU calculations provide the best overall description of the experimental data along the Ar isotopic chain including the $\emph{E}$(2$_{1}^{+}$) of $^{52}$Ar. The VS-IMSRG approach using the 1.8/2.0 (EM) interaction reasonably reproduces the measured $\emph{E}$(2$_{1}^{+}$) in neutron-rich Ar isotopes, though an overprediction is seen between $\emph{N}$ = 28 and 34.
The dependence of the $\emph{ab initio}$ calculations on the initial $\emph{NN}$ and 3$\emph{N}$ forces is illustrated by the VS-IMSRG calculations with the N$^{2}$LO$\rm_{sat}$ Hamiltonian. Compared to results with the 1.8/2.0~(EM) interaction, the N$^{2}$LO$\rm_{sat}$ Hamiltonian systematically underpredicts the data, despite a better agreement at $\emph{N}$ = 28 and 30.
The DCE-EOM calculations with the 1.8/2.0~(EM) interaction reproduce the $\emph{E}$(2$_{1}^{+}$) in $^{44,48}$Ar within the estimated uncertainties, but underestimate the $\emph{E}$(2$_{1}^{+}$) in $^{52}$Ar by $\sim$600 keV. The 2PR-EOM result for $^{52}$Ar is consistent with the DCE-EOM calculation, but fails to reproduce the steep increase of $\emph{E}$(2$_{1}^{+}$) from $^{50}$Ar to $^{52}$Ar. We note that 2PR-EOM gives a $\emph{E}$(2$_{1}^{+}$) at 3.0~MeV for $^{46}$Ar, consistent with the $N=28$ shell closure but almost twice the experimental value. For $^{40}$Ar, which is characterized by deformation and shape co-existence \cite{bit83}, all considered calculations underestimate its $\emph{E}$(2$_{1}^{+}$).

Despite being rooted in the same chiral effective interaction, namely, 1.8/2.0 (EM), the coupled-cluster and VS-IMSRG approaches predict very different behaviours about the change of $\emph{E}$(2$_{1}^{+}$) from $^{50}$Ar to $^{52}$Ar. However, for closed (sub-)shell Ca isotopes, theses two calculations give consistent results. The differences in calculated $\emph{E}$(2$_{1}^{+}$) in neutron-rich Ar isotopes indicate that the total theoretical uncertainties might be larger than the estimated error bars shown in Fig. 3. The observed steep rise of $\emph{E}$(2$_{1}^{+}$) from $^{50}$Ar and $^{52}$Ar serves as an important benchmark to understand the uncertainties of the employed many-body methods.

It is worth noting that the modified SDPF-MU shell model calculations and the VS-IMSRG approach using the 1.8/2.0 (EM) interaction have both been used along the $\emph{N}$~=~34 isotonic chain to investigate the shell evolution. Both calculations suggest that the $\emph{N}$~=~34 shell gap persists from $^{54}$Ca towards more exotic $\emph{N}$~=~34 isotones, which is consistent with the measured high-lying 2$_{1}^{+}$ state in $^{52}$Ar presented here. However, we would like to emphasize that there is no direct correlation between the measured $\emph{E}$(2$_{1}^{+}$) and the size of the shell gap which is defined as the difference between the effective single-particle energies (ESPEs), since the latter is not an observable. Indeed, calculations predicting similar $\emph{E}$(2$_{1}^{+}$) might give different magnitudes of shell gaps. As it is the case here, the ESPEs extracted by the modified SDPF-MU calculations using the method described in Ref. \cite{utsuno99} indicate that the $\emph{N}$ = 34 shell gap in $^{52}$Ar ($\sim$3.1 MeV) exceeds that in $^{54}$Ca ($\sim$2.6 MeV) \cite{step15}. Conversely, the ESPEs calculated by the VS-IMSRG approach using the method of Ref.~\cite{dug15} suggest the $\emph{N}$ = 34 shell gap in $^{52}$Ar ($\sim$2.6 MeV) is smaller than that in $^{54}$Ca ($\sim$3.2 MeV).
In addition, the VS-IMSRG approach also provides the orbital occupancies of the 0$_{1}^{+}$ and 2$_{1}^{+}$ states in $^{52}$Ar and $^{54}$Ca. It reveals that in the 2$_{1}^{+}$ excitation of $^{52}$Ar only $\sim$0.5 neutrons are excited from $\emph{p}_{1/2}$ to $\emph{f}_{5/2}$ and proton excitations also contribute due to the open proton shell, whereas in the case of $^{54}$Ca, $\sim$0.9 neutrons are excited across the $\emph{N}$ = 34 shell gap. This is consistent with the observed decrease in $\emph{E}$(2$_{1}^{+}$) between $^{54}$Ca and $^{52}$Ar.
Nevertheless, both calculations predict $^{48}$Si as a new doubly magic nucleus. The $\emph{E}$(2$_{1}^{+}$) of $^{48}$Si in SDPF-MU \cite{utsu14} and VS-IMSRG \cite{jason18} calculations lies at 2.85 and 3.13 MeV, respectively. However, it is not yet known whether $^{48}$Si ground state and its 2$_{1}^{+}$ state are stable against neutron emission. Mass models that reproduce well the observed limits of existence in the $\emph{pf}$-shell region \cite{tar18} tend to predict $^{48}$Si as a drip-line nucleus in which continuum effects might also play an important role in the structure of $^{48}$Si.

\raggedbottom
To summarize, the low-lying structure of $^{52}$Ar was investigated using the $^{53}$K(\emph{p},2\emph{p})$^{52}$Ar reaction at $\sim$210 MeV/u. The measured 2$^{+}_{1}$ state lies at 1656(18)~keV, the highest among the Ar isotopes with $\emph{N}$ $>$ 20. The measured ($\emph{p}$,2$\emph{p}$) cross sections to different final states of $^{52}$Ar agree with calculations and support the proposed spin-parity assignments. Shell-model calculations with phenomenological and the chiral interaction 1.8/2.0~(EM) both reproduce the measured 2$^{+}_{1}$ systematics of the neutron-rich Ar isotopes, and suggest a $\emph{N}$~=~34 subshell closure in $^{52}$Ar. However, coupled-cluster calculations based on the same chiral interaction underestimate the 2$^{+}_{1}$ excitations in $^{52}$Ar. The measured $\emph{E}$(2$_{1}^{+}$) of $^{52}$Ar serves as an important benchmark to understand the uncertainties of the employed many-body methods. Our results offer the first experimental signature of the persistence of the $\emph{N}$ = 34 subshell closure below $\emph{Z}$ = 20, and agree with shell-model calculations predicting $^{48}$Si as a new doubly magic nucleus far from stability.\\[12pt]
We thank the RIKEN Nishina Center accelerator staff for their work in the primary beam delivery and the BigRIPS team for preparing the secondary beams. We thank Y.~Utsuno for providing us the SDPF-MU calculated 2$^{+}_{1}$ excitation energies for Ar isotopes and $^{48}$Si, and A. Poves for providing us the SDPF-U calculated 2$^{+}_{1}$ excitation energies for Ar isotopes. The development of MINOS has been supported by the European Research Council through the ERC Grant No. MINOS$-$258567. H.~N.~L. thanks G. Schnabel and R. Taniuchi for valuable discussions about the significance level estimation and acknowledges the support from the Enhanced Eurotalents program (PCOFUND-GA-2013-600382) co-funded by CEA and the European Union.
H.~N.~L., A. O. and A. S. acknowledge the support from the Deutsche Forschungsgemeinschaft (DFG, German Research Foundation) $-$ Projektnummer 279384907 $-$ SFB 1245.
C.~A.~B. acknowledges support from the U.S. NSF grant No. 1415656 and the U.S. DOE grant No. DE-FG02-08ER41533.
J. D. H. and R. S. acknowledge the support from NSERC and the National Research Council Canada.
Y.~L.~S. acknowledges the support of Marie Sk\l{}odowska-Curie Individual Fellowship (H2020-MSCA-IF-2015-705023) from the European Union. 
I.~G. has been supported by HIC for FAIR and Croatian Science Foundation. 
L.~X.~C. and B.~D.~L. have been supported by the Vietnam MOST through the Physics Development Program Grant No.\DJ T\DJ LCN.25/18. K.~I.~H., D.~K. and S.~Y.~P. have been supported by the NRF grant funded by the Korea government (No. 2017R1A2B2012382 and 2018R1A5A1025563).
This work was supported in part by JSPS KAKENHI Grant No. 16H02179, MEXT KAKENHI Grants No. 24105005 and No. 18H05404.
This work was also supported by the Office of Nuclear Physics, U.S. Department of Energy, under grants de-sc0018223 (NUCLEI SciDAC-4 collaboration) and the Field Work Proposal ERKBP72 at Oak Ridge National Laboratory (ORNL). Computer time was provided by the Innovative and Novel Computational Impact on Theory and Experiment (INCITE) program. This research used resources of the Oak Ridge Leadership Computing Facility located at ORNL, which is supported by the Office of Science of the Department of Energy under Contract No.  DE-AC05-00OR22725. 

\raggedbottom

\begin{thebibliography}{99}
\bibitem{may55} M. Mayer and J. H. D. Jensen, Elementary Theory of Nuclear Shell Structure (Wiley, New York, 1955).
\bibitem{sorlin08} O. Sorlin and M.-G. Porquet, Prog. Part. Nucl. Phys. {\bf 61}, 602 (2008).
\bibitem{thibault75} C. Thibault \emph{et al.}, Phys. Rev. C {\bf 12}, 644 (1975).
\bibitem{mue84} D. Guillemaud-Mueller \emph{et al.},  Nucl. Phys. A {\bf 426}, 37 (1984). 
\bibitem{ba07} B. Bastin \emph{et al.}, Phys. Rev. Lett. {\bf 99}, 022503 (2007).
\bibitem{ozawa00} A. Ozawa, T. Kobayashi, T. Suzuki, K. Yoshida and I. Tanihata, Phys. Rev. Lett. {\bf 84}, 5493 (2000).
\bibitem{tshoo12} K. Tshoo \emph{et al.}, Phys. Rev. Lett. {\bf 109}, 022501 (2012).
\bibitem{tal60} I. Talmi and I. Unna, Phys. Rev. Lett. {\bf 4}, 469 (1960).
\bibitem{ots05} T. Otsuka, T. Suzuki, R. Fujimoto, H. Grawe, and Y. Akaishi, Phys. Rev. Lett. {\bf 95}, 232502 (2005).
\bibitem{zuker03} A. P. Zuker, Phys. Rev. Lett. {\bf 90}, 042502 (2003).
\bibitem{ots10} T. Otsuka \emph{et al.}, Phys. Rev. Lett. {\bf 104}, 012501 (2010).
\bibitem{bohr88} A. Bohr and B.R. Mottelson, $\emph{Nuclear Structure}$, Vol. 1 (World Scientific, Singapore, 1998).
\bibitem{huck85}  A. Huck \emph{et al.}, Phys. Rev. C {\bf 31}, 2226 (1985).
\bibitem{jan02} R. V. F. Janssens \emph{et al.}, Phys. Lett. B {\bf 546}, 55 (2002).
\bibitem{pri01} J. I. Prisciandaro \emph{et al.}, Phys. Lett. B {\bf 510}, 17 (2001).
\bibitem{step15} D. Steppenbeck \emph{et al.}, Phys. Rev. Lett. {\bf 114}, 252501 (2015).
\bibitem{din05} D.-C. Dinca \emph{et al.}, Phys. Rev. C {\bf 71}, 041302(R) (2005).
\bibitem{bur05} A. B\"{u}rger \emph{et al.}, Phys. Lett. B {\bf 622}, 29 (2005).
\bibitem{wie13} F. Wienholtz \emph{et al.}, Nature (London) {\bf 498}, 346 (2013).
\bibitem{ro15} M. Rosenbusch \emph{et al.}, Phys. Rev. Lett. {\bf 114}, 202501 (2015).
\bibitem{xu15} X. Xu \emph{et al.}, Chin. Phys. C {\bf 39}, 104001 (2015).  
\bibitem{rui16} R. F. Garcia Ruiz \emph{et al.} Nature (London) {\bf 12}, 594 (2016).
\bibitem{lei18} E. Leistenschneider \emph{et al.}, Phys. Rev. Lett. {\bf 120}, 062503 (2018).
\bibitem{step13} D. Steppenbeck \emph{et al.}, Nature (London) {\bf 502}, 207 (2013).
\bibitem{mi18} S. Michimasa \emph{et al.}, Phys. Rev. Lett. {\bf 121}, 022506 (2018).
\bibitem{su13} H. Suzuki \emph{et al.}, Phys. Rev. C {\bf 88}, 024326 (2013).
\bibitem{zhu06} S. Zhu \emph{et al.}, Phys. Rev. C {\bf 74}, 064315 (2006).
\bibitem{step17} D. Steppenbeck \emph{et al.}, Phys. Rev. C {\bf 96}, 064310 (2017).
\bibitem{hagen12} G. Hagen, M. Hjorth-Jensen, G. R. Jansen, R. Machleidt, and T. Papenbrock, Phys. Rev. Lett. {\bf 109}, 032502 (2012).
\bibitem{utsu14} Y. Utsuno \emph{et al.}, JPS Conf. Proc. {\bf 6}, 010007 (2015).
\bibitem{fukuda13} N. Fukuda, T. Kubo, T. Ohnishi, N. Inabe, H. Takeda, D. Kameda, and H. Suzuki, Nucl. Instrum. Methods Phys.
Res., Sect. B {\bf 317}, 323 (2013).
\bibitem{ku12} T. Kubo \emph{et al}., Prog. Theor. Exp. Phys. {\bf 2012}, 03C003 (2012).
\bibitem{ko13} T. Kobayashi \emph{et al}., Nucl. Instrum. Methods Phys. Res., Sec. B {\bf 317}, 294 (2013).
\bibitem{ob14} A. Obertelli \emph{et al.}, Eur. Phys. J. A {\bf 50}, 8 (2014). 
\bibitem{san18} C. Santamaria \emph{et al}., Nucl. Instrum. Methods Phys. Res., Sec. A {\bf 905}, 138 (2018).
\bibitem{ta14} S. Takeuchi \emph{et al}., Nucl. Instrum. Methods Phys. Res., Sec. A {\bf 763}, 596 (2014).
\bibitem{ago03} S. Agostinelli \emph{et al.}, Nucl. Instrum. Methods Phys. Res., Sect. A {\bf 506}, 250 (2003).
\bibitem{raman01} S. Raman \emph{et al.}, At. Data Nucl. Data Tables {\bf 78}, 1 (2001).  
\bibitem{wang17} M. Wang, G. Audi, F. G. Kondev, W. J. Huang, S. Naimi, and X. Xu, Chin. Phys. C {\bf 41}, 030003 (2017).
\bibitem{nndc} http://www.nndc.bnl.gov/.
\bibitem{nowak16} K. Nowak\emph{et al.}, Phys. Rev. C {\bf 93}, 044335 (2016).
\bibitem{tsu12} K. Tsukiyama, S. K. Bogner, and A. Schwenk, Phys. Rev. C {\bf 85}, 061304(R) (2012).
\bibitem{str16} S. R. Stroberg, H. Hergert, J. D. Holt, S. K. Bogner, and A. Schwenk, Phys. Rev. C {\bf 93}, 051301(R) (2016).
\bibitem{her16} H. Hergert, S. K. Bogner, T. D. Morris, A. Schwenk, and K. Tsukiyama, Phys. Rep. {\bf 621}, 165 (2016).
\bibitem{str17} S.~R.~Stroberg \emph{et al.}, Phys. Rev. Lett. {\bf 118}, 032502 (2017).
\bibitem{sim17} J. Simonis, S. R. Stroberg, K. Hebeler, J. D. Holt, and A. Schwenk, Phys. Rev. C {\bf 96}, 014303 (2017).
\bibitem{bar07} R.J. Bartlett and M. Musia\l, Rev. Mod. Phys. {\bf 79}, 291 (2007).
\bibitem{hagen14} G. Hagen, T. Papenbrock, M. Hjorth-Jensen, and D. J. Dean, Rep. Prog. Phys. {\bf 77}, 096302 (2014).
\bibitem{epe09} E. Epelbaum, H.-W. Hammer, and U.-G. Mei$\ss$ner, Rev. Mod. Phys. {\bf 81}, 1773 (2009).
\bibitem{mac11} R. Machleidt and D. R. Entem, Phys. Rep. {\bf 503}, 1 (2011).
\bibitem{heb11} K. Hebeler, S. K. Bogner, R. J. Furnstahl, A. Nogga, and A. Schwenk, Phys. Rev. C {\bf 83}, 031301 (2011).
\bibitem{sim16} J. Simonis, K. Hebeler, J.D. Holt, J. Men\'{e}ndez, and A. Schwenk, Phys. Rev. C {\bf 93}, 011302(R) (2016).
\bibitem{eks15} A. Ekstr\"{o}m \emph{et al.}, Phys. Rev. C {\bf 91}, 051301(R) (2015).
\bibitem{jansen2011} G. R. Jansen, M. Hjorth-Jensen, G. Hagen, and T. Papenbrock, Phys. Rev. C {\bf 83}, 054306 (2011).
\bibitem{watts1996} John D. Watts and Rodney J. Bartlett, Chemical Physics Letters {\bf 258}, 581 (1996).
\bibitem{nowacki09} F. Nowacki and A. Poves, Phys. Rev. C {\bf 79}, 014310 (2009).
\bibitem{utsu12} Y. Utsuno, T. Otsuka, B. A. Brown, M. Honma, T. Mizusaki, and N. Shimizu, Phys. Rev. C {\bf 86}, 051301(R) (2012).
\bibitem{pap13} J. Papuga \emph{et al.}, Phys. Rev. Lett. {\bf 110}, 172503 (2013).
\bibitem{auman13} T. Aumann, C. A. Bertulani, and J. Ryckebusch, Phys. Rev. C {\bf 88}, 064610 (2013).
\bibitem{ber10} C. A. Bertulani and C. De Conti, Phys. Rev. C {\bf 81}, 064603 (2010).
\bibitem{lap93} L. Lapik\'as, Nucl. Phys. A {\bf 553}, 297 (1993).
\bibitem{atar18} L. Atar \emph{et al.}, Phys. Rev. Lett. {\bf 120}, 052501 (2018).
\bibitem{kawase18} S. Kawase \emph{et al.}, Prog. Theor. Exp. Phys. {\bf 2018}, 021D01 (2018).
\bibitem{bit83} E. Bitterwolf \emph{et al.}, Z. Physik A {\bf 313}, 123 (1983). 
\bibitem{utsuno99} Y. Utsuno, T. Otsuka, T. Mizusaki and M. Honma, Phys. Rev. C \textbf{60}, 054315 (1999).
\bibitem{dug15} T. Duguet, H. Hergert, J. D. Holt, and V. Soma, Phys. Rev. C \textbf{92}, 034313 (2015).
\bibitem{jason18} J. D. Holt \emph{et al.} (manuscript in preparation).
\bibitem{tar18} O. B. Tarasov \emph{et al.}, Phys. Rev. Lett. {\bf 121}, 022501 (2018).

\end{thebibliography}
%

\end{document}